\documentclass[a4paper]{article}
\usepackage{multirow}
\usepackage{url}
\usepackage{xcolor}

\usepackage{INTERSPEECH2021}

\title{Revisiting Speech Content Privacy}
\name{Jennifer Williams $^1$, Junichi Yamagishi $^2$, Paul-Gauthier No\'e $^3$, Cassia Valentini-Botinhao $^1$, Jean-Fran\c{c}ois Bonastre$^3$}
\address{
  $^1$Centre for Speech Technology Research, University of Edinburgh, UK\\
  $^2$National Institute for Informatics, Japan\\
  $^3$Laboratoire Informatique d’Avignon (LIA), Avignon Universit\'e, France}

\email{j.williams@ed.ac.uk}

\begin{document}

\maketitle
\begin{abstract}
In this paper, we discuss an important aspect of speech privacy: \textit{protecting spoken content}. New capabilities from the field of machine learning provide a unique and timely opportunity to revisit speech content protection. There are many different applications of content privacy, even though this area has been under-explored in speech technology research. This paper presents several scenarios that indicate a need for speech content privacy even as the specific techniques to achieve content privacy may necessarily vary. Our discussion includes several different types of content privacy including recoverable and non-recoverable content. Finally, we introduce evaluation strategies as well as describe some of the difficulties that may be encountered.  

\end{abstract}
\noindent\textbf{Index Terms}: privacy, speech coding, speech recognition

\section{Introduction}

Speech content privacy refers to the ability to conceal or mask sensitive content information within the speech signal. Determining what would be considered sensitive information ultimately depends on the use-cases. Private content may be found within particular keywords or keyphrases such as named entities (places, dates, locations, organizations etc.), or financial and medical details. In this paper, we use the term \textit{speech content} to mean semantically meaningful words. However, this definition could be reasonably extended to paralinguistic information including mannerisms, patterns of disfluency, or high-level perceptual features like prosody and emotion.

Traditionally, speech content privacy has been rooted in the idea that a signal-emitting device can be set up within a physical space, such as an office room, to conceal what people say during private conversations. This device emits a special type of noise to mask semantically-relevant speech sounds, like words or phonemes~\cite{akagi2012privacy, kondo2014gender, donley2016improving}. Such approaches can effectively mask speech content to the point of rendering it unintelligible to nearby eavesdroppers. However, this approach provides a blanket solution. It is heavily dependent on specific room and speaker characteristics, making it challenging for the technology to generalize to a variety of scenarios~\cite{phunruangsakao2020speech}. 

Furthermore, it may not always be desireable to mask entire conversations but instead only specific content that would be considered sensitive in nature. Perhaps the most well-known form of content privacy comes from broadcasting where a sensitive phrase is masked or replaced using a `bleep' sound. There are downsides to using a `bleep': it does not preserve speaker characteristics, it interrupts the listening experience, and it may be irreversible. In fact, it may be preferable to simply replace a sensitive phrase with a less-sensitive counterpart. If there were perfect text-to-speech (TTS) synthesis and perfect automatic speech recognition (ASR), we could find a sensitive phrase automatically via ASR and we could replace it with a less-sensitive phrase, while maintaining characteristics of the original speakers' voice and style. The current deep learning approaches for ASR and TTS are not ideal, but, still it is possible to produce such results with these applications\footnote{\url{https://www.descript.com/}}. 


Although there has been recent scientific focus on speaker-centric privacy and security~\cite{nautsch2019preserving}, now there is an increasing need to expand speech privacy to ensure that \textit{spoken content} is also protected. Solutions to speech content privacy have not yet been fully explored because new use-cases are still emerging. While more and more people adapt to voice-based technologies, two main privacy issues have become prominent. First, many people have a reasonable expectation of privacy when it comes to how their devices store, process, and transmit their voice data~\cite{blumenthal2009multiple}. Second, some people modify their personal behavior due to privacy concerns, such as never using voice-enabled devices in open or public spaces~\cite{vimalkumar2021okay}. Both of these issues must be addressed in order for voice technology to reach full potential.
 
During the past few years, there were significant advances in machine learning, especially for deep neural networks (DNNs). This has been transformational to the speech technology landscape. Because of DNNs, it is possible to train models using federated learning wherein models are adapted to specific user data without the need to transmit data away from a personal device~\cite{yang2019federated}. DNNs have also enabled the development of neural vocoders that produce extremely high-quality synthetic speech~\cite{vanwavenet, kalchbrenner2018efficient} as well as multiple different approaches to speech signal disentanglement that separates speech content from speaker identity~\cite{ebbers2021contrastive, williams2021learning}. This paper explores how new forms of content privacy can be developed with different use-cases in mind and provides an opportunity to re-imagine how content privacy can be used to meet the needs of society.

\section{Recent Work}
There is an apparent trade-off between content-based privacy and the ability to use speech in downstream tasks. Recent work in \cite{turan2021adapting} examined how privacy-transformed data affects the ability to train models for automatic speech recognition (ASR) by masking named entities in the \text{text} data during model training. In particular, they sought to adapt existing text language models to account for missing (or `masked') words from the data. While content privacy initially reduced overall ASR system performance, they successfully developed a method to adapt text language models and regained some performance. They did not describe any methods for masking the speech audio. 

The Bavarian Archive for Speech Signals (BAS) offers an online webservice\footnote{\url{https://clarin.phonetik.uni-muenchen.de/BASWebServices/interface}} that will mask speech content using a pipeline approach. An audio file is submitted to the webservice along with a list of target words. ASR is used to obtain forced alignments of words and timestamps from the audio file. The content is masked with white noise, silence, or a bleep, and a new audio file is produced. While this pipeline approach could be useful for static databases, the required forced alignments makes this solution too computationally expensive to extend to privacy scenarios that require real-time performance.

Another aspect of content privacy is related to speech codecs and compression. A new generative DNN architecture was introduced in \cite{casebeer2021enhancing} and \cite{williams2021exploring}, independently, with different speech technologies in mind. The architecture is a dual-encoder vector quantized variational autoencoder (VQ-VAE) that learns to disentangle speech content and speaker identity information in the speech signal while simultaneously creating a discrete and compressed representation of the speech. In \cite{casebeer2021enhancing} the goal was to use VQ-VAE to compress the speech signal and enhance it by removing unwanted noise. They measured compression rate as bits/sec as well as human judgements of the enhanced speech naturalness. In \cite{williams2021exploring}, they took advantage of the discrete representations to \textit{mask the content} of targeted phrases using different types of masks. They measured the resulting intelligibility and human judgements of speaker consistency. Taken together, work on this VQ-VAE architecture is promising for content-based privacy because it effectively separates content from other information in the speech signal. The VQ-VAE design is also useful for the privacy scenario that involves speech compression and transmission, discussed in the next section. 

\section{Content Privacy Scenarios}
Content privacy extends to any situation where sensitive information is delivered using voice. In this section, we outline four prominent scenarios that are timely and relevant given the current state of the art for speech technology. There may be some overlap between the privacy needs of the scenarios, however it is possible that the technical solutions will vary.

\subsection{Voice Storage Privacy}
All voice-enabled devices capture and store speech data, and in some cases also transmit it from the device to a larger database on a remote computing server. Speech capture and storage will sometimes involve private conversations as a result of intentional or unintentional recordings. While some storage mechanisms have security measures by design, such as secure enclaves on mobile devices or data encryption, content-based privacy offers an additional layer of protection~\cite{qian2018towards}. Numerous functionalities for voice-enabled devices require the ability to access stored voice content. Speech recognition on mobile devices is one example that requires storing speech data, though it is possible to fully encrypt all speech content while performing speech recognition~\cite{glackin2017privacy}. In addition, it would be beneficial to create ``edge'' privacy solutions that can mask sensitive content on a device when the speech is first captured by the microphone, preferably also within a secure enclave. Successful content privacy could enable researchers to utilize speech databases for research and development that would otherwise be prohibited, due to legal issues with the European Union GDPR \cite{nautsch2019gdpr}.

\subsection{Speech Compression, Transmission, and Broadcast}
In order to transmit or broadcast speech, it must undergo compression which creates a more compact representation of the data. This is true for mobile phones, internet voice calling, and television broadcast, among others. It is possible for an intruder to intercept, eavesdrop and even maliciously manipulate speech content during transmission. Watermarking has been proposed as a countermeasure solution~\cite{chen2007content}. However, watermarking only helps to ensure the message remains unchanged and it does not conceal sensitive content. Other countermeasures such as voice scramblers would require highly specialized hardware and software solutions, and some intelligibility may be lost during the scrambling and unscrambling process~\cite{de2008speech}. Related broadcast scenarios include law court testimony, emergency calls, and policing: it is important to protect the witness identity, which implies hiding or masking different levels of information, from voice identity to linguistic content. At the same time, it is important to retain other non-sensitive information as much as possible in order to provide valid testimony. Sensitive testimony may also require redaction before being released to the public or while it is being televised~\cite{roberts1998empirical}.

\subsection{Speech and Speaker Recognition}
\label{sec:speechspeakerreco}
As mentioned earlier, there is a trade-off between protecting sensitive content while also using the speech for downstream tasks such as ASR and speaker recognition. Certain speech technologies may require speaker information to remain unaltered by the content masking. Authentication by voice, also known as automatic speaker verification (ASV)~\cite{bimbot2004tutorial}, is a common application where the speaker information needs to be preserved when using content masking at the same time. ASV typically compares two utterances (a \emph{reference} with a \emph{test} utterance) to produce a score, subject to binary decision. The score is high if the two utterances likely come from the same speaker (\emph{target} proposition) or the score is low if the utterances come from different speakers (\emph{impostor} proposition). A related application is speaker diarization~\cite{spkdia} which aims to detect \emph{who spoke when?} in a multi-speaker conversation. Content masking can be used in this scenario if it does not alter speaker information or cause confusion between speakers. 

\subsection{Voice-Enabled Assistive Technology}
When a user interacts with their voice assistant, such as Siri or Google, the responses spoken aloud from the device may contain sensitive information that the user does not want people nearby to overhear~\cite{vimalkumar2021okay,easwara2015privacy,liao2019understanding}. Blind users of speech technology face unique challenges for privacy when using the internet and they are some of the heaviest users of text-to-speech (TTS) synthesis since it is the core technology of screen-readers~\cite{regal2016insights}. While sighted people can enjoy a particular amount of content privacy from quietly reading text on a screen, this type of privacy is not extended to blind users who must have screens read aloud using screen-reading software~\cite{abdolrahmani2018siri}. Focus groups show that in public spaces blind users may hold a device very close to their ear or use earbuds, even though this practice is potentially hazardous because it blocks out other environmental noises that blind users must attend to~\cite{47021}. There are many nuances to developing speech technologies that are optimized for blind users. Content privacy is under-explored and could have a big impact for accessibility~\cite{kuriakose2020tools}. It may be possible to embed privacy capabilities within the TTS screen-reader, but that would prevent the user from receiving important information. Another possibility is to develop a special earbud customized for blind users, while allowing for other important sounds in the environment.




\section{Privacy Approaches}
\begin{figure}
\centering
     \includegraphics[width=1.0\linewidth]{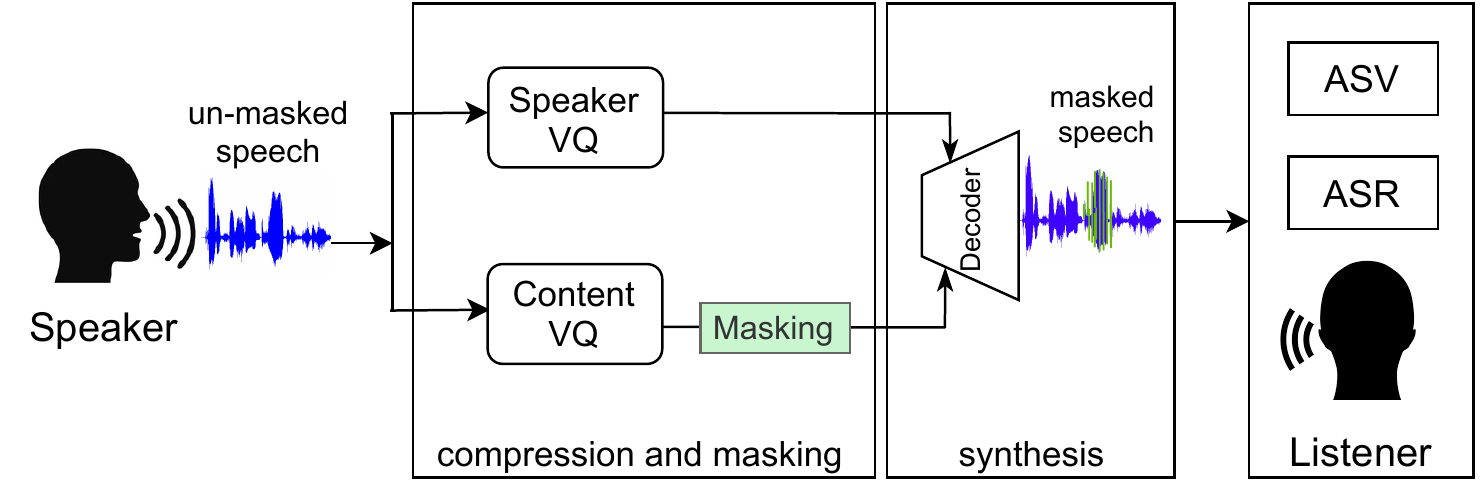}
     \caption{Simultaneous speech compression and content masking using a deep learning technique called VQ-VAE.}
\label{fig:masking}
\end{figure}

Ideally, the solutions to content privacy would be optimized for the specific needs and requirements of each particular scenario. For example, a customized earbud for blind users of screen-readers can be very different from privacy solutions for speech databases. Even still, as far as speech is concerned there are some general approaches that are worth discussing as a starting point. Speech signal disentanglement stands out as a promising overall approach. Disentanglement is a form of distributed representation learning that separates different types of speech information into separate representations, such as speaker identity and speech content. One of the main benefits of disentanglement is that it allows content to be modified separately from other informational factors. In addition, some disentanglement techniques such as VQ-VAE also compress the speech signal, which is beneficial for transmission scenarios~\cite{casebeer2021enhancing, williams2021exploring}. The VQ-VAE approach is shown in Figure~\ref{fig:masking}. Original, un-masked speech is disentangled into separate representations of speaker identity and content using two vector-quantized (VQ) codebooks. These codebooks are highly compressed The content VQ codebook is used for content masking. The masked speech is synthesized, and made available to human listeners or downstream speech technologies.

Speech content can be masked through various mechanisms and one of the most obvious would be to replace sensitive content with silence or noise. However, there have not yet been any studies to determine how different types of masks impact performance on downstream speech tasks such as ASR, ASV, or human listening effort and intelligibility. Another issue is whether or not a content mask would be reversible in the sense that the speech content could be concealed and later recovered. A reversible privacy mask might not be ideal for speech databases described in Section 3.1 if the speech data will be shared with third-parties. On the other hand, a reversible mask would be useful for speech transmission described in Section 3.2 where the goal may be to protect content from being intercepted only during transmission.

\section{Content Privacy Evaluation}
Currently there are no established protocols for evaluating speech content privacy. While it is not the goal of this paper to definitively specify an evaluation protocol, we present three assessment viewpoints that can help inform future efforts: task-based, high-level, and low-level. 

\subsection{Task-Based Assessment}
When applying ASV on content-masked speech it is important to ensure that speaker information is not altered. The speech output of a content masking system may not be perfect due to compression or the quality of speech vocoder. The result may be a speaker identity shift in the synthetic voice space, even for portions of speech without content masking such as surrounding words. Consequently, a natural and a content-masked utterance coming from the same speaker could be marked as two different speakers. One way to assess if content-masking has affected the speaker identity is to compare un-masked enrolment utterances with masked test utterances. Whereas to check speaker separability in the protected space, the enrolment and test utterances would both be masked. As explained in Section~\ref{sec:speechspeakerreco}, an ASV system compares an enrolment and test utterance, and produces a similarity score. Then a threshold, also known as the operating point, is set in order to decide between the target and impostor propositions. There are two kinds of possible errors for the ASV system: false alarms (FA) and false rejections (FR). Several metrics are used to interpret these errors. The equal error rate (EER) describes the operating point at which the FA and the FR rates are equal. The log-likelihood-ratio cost~\cite{brummercllr} is also commonly used to evaluate ASV systems. It measures the ability of the scores to resemble calibrated log-likelihood-ratios (LLRs) and is thus independent of the operating point. 


Content-masking can also be evaluated in terms of an ASR-style task, especially for privacy scenarios that require high intelligibility for the un-masked and non-sensitive surrounding words. A word error rate (WER) or phone error rate (PER) measures how many words or phones are correctly recognised by an ASR system. The WER/PER should decline proportionally with the quantity of words that have been masked. Human judgements of intelligibility have similar requirements. One way to assess the impacts on ASR and intelligibility is to present human listeners with masked speech audio alongside ASR output and measure how often listeners agree with the ASR transcript based on what they hear~\cite{turan2021adapting}. This hybrid style of aligning human judgements with task-based performance holds for ASR as well as ASV and speaker separation.

\begin{figure}
\centering
     \includegraphics[width=0.85\linewidth]{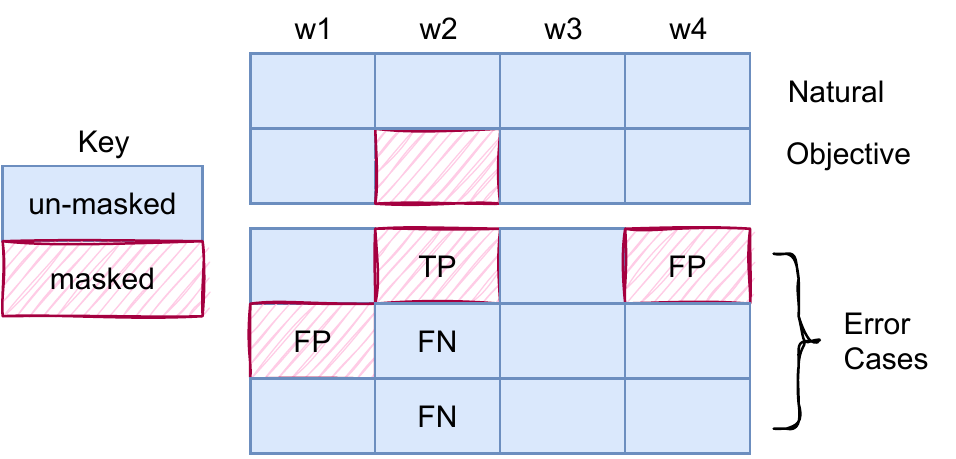}
     \caption{Sample word-level masking errors and decisions which can be compared to natural (un-masked) and objective (with masking), where $w_1, w_2.. w_n$ is a sequence of words.}
\label{fig:mer}
\end{figure}

\begin{figure*}[ht!]
\centering
     \includegraphics[width=0.86\linewidth]{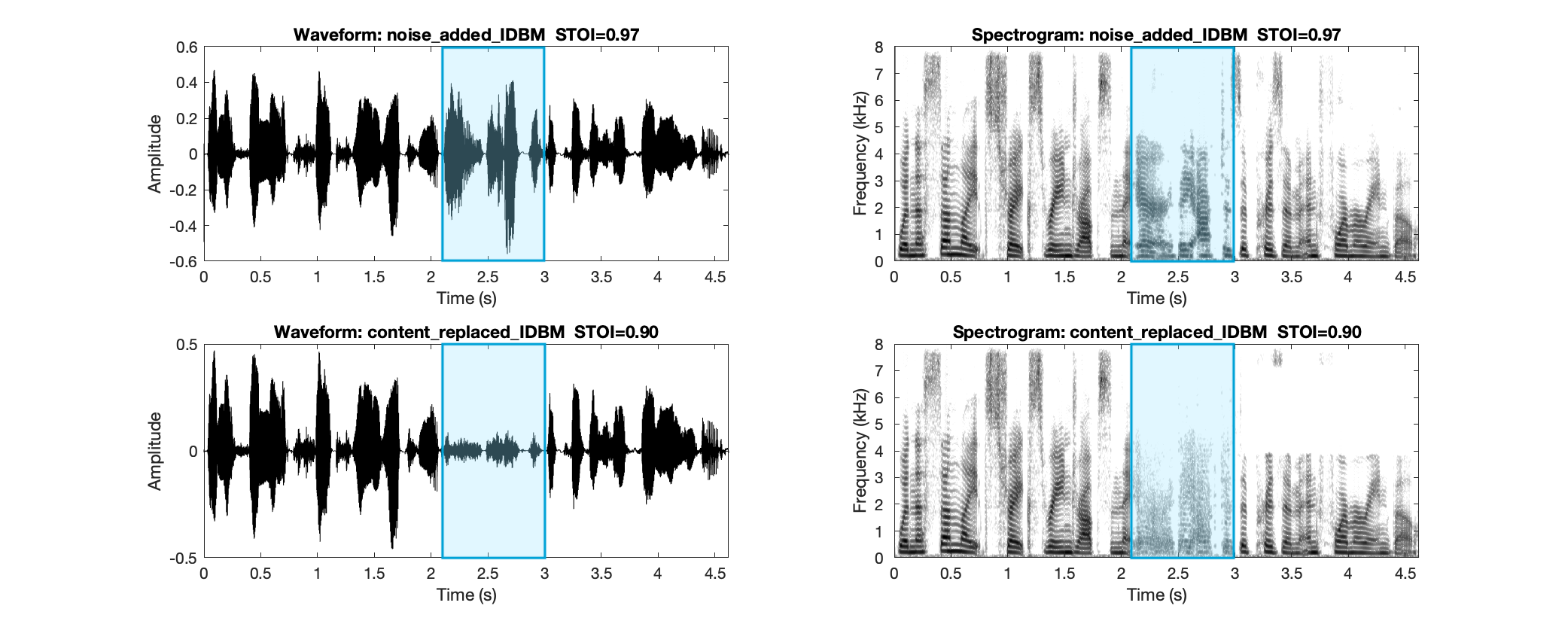}
     \caption{Waveform and spectrogram to compare ideal binary mask from two different content-masking approaches: additive noise (top) and replacement with noise (bottom). In each case, target content to be masked is highlighted by the box. \\``When sunlight strikes raindrops in the \textbf{[air they act as a]} prism and form a rainbow.''}
\label{fig:IDBM}
\end{figure*} 

\subsection{High-Level Assessment}
This paper has been discussing content privacy based on the notion that \textit{content} refers to spoken words. For speech privacy and security, different types of word mask errors imply different consequences. For example, failing to mask a target sensitive phrase might invoke a penalty whereas accidentally masking a non-sensitive phrase would not. As discussed earlier, some privacy scenarios require that non-sensitive words remain highly intelligible for human listeners or for other speech technologies like ASR. We introduce a metric called \textit{mask error rate} (MER) which can be used specifically to assess errors about which words have been masked correctly. Consider four types of errors/decisions (Figure~\ref{fig:mer}) at the word-level: True negative (TN) word is correctly unmasked; True positive (TP) word is correctly masked; False negative (FN) word is incorrectly unmasked; False positive (FP) word is incorrectly masked.

The MER metric can be described by Equation~\ref{eq:mer}. For a given utterance, the weighted combination of masking errors with respect to the length of the utterance: 

 \begin{equation} 
MER_{utt} = \frac{\alpha FN + \beta FP}{TN + TP + FN + FP}
 \label{eq:mer}
 \end{equation}

where $\alpha$ and $\beta$ are penalties that can be used to balance the two types of word masking errors. For speech privacy, an FN error may need to be weighted more heavily because it could lead to revealing sensitive information. This metric could be adapted for any size of chunk, larger or smaller than the word level. One potential limitation is that MER requires high-quality time alignments in order to perform the calculations. To be useful in practice, MER should also account for slight variation of word boundaries, such as inadvertently causing surrounding words to become unintelligible. For reversible masking, MER can be adapted to measure how much content is recoverable when the mask is reversed.


\subsection{Low-Level Assessment}

It may also be possible to develop an evaluation metric borrowing from a technique in speech enhancement, called \textit{ideal binary masks}. An idea binary mask is used to remove various types of noise from the speech signal (babble, static, reverberation, etc). There are many different versions of this technique. Overall it can be summarized as comparing the original clean signal with a noisy corrupted signal, and computing signal-to-noise ratio (SNR) to identify which areas of the speech signal could be attenuated. This effectively removes the noise portions of the signal, while leaving the speech portions of the signal intact. An ideal binary mask describes the perfect (\textit{idealized}) solution of removing noise in the speech signal so that the noise (and only the noise) is completely removed, while also preserving speech intelligibility~\cite{loizou2010reasons}.

Consider two versions of speech processed with an ideal binary mask shown as waveforms and spectrograms in Figure~3~\cite{IDBM}. The top was created by \textit{adding} a temporally-modulated speech-shaped noise masker (ICRA noise 9 from~\cite{cooke2013evaluating}) to the signal, to mask a target phrase, and then an ideal binary mask was calculated and applied to attempt to recover the speech. The bottom was created by \textit{replacing} a target phrase with the same type of noise. Recovery of the target phrase can be measured by a short term objective intelligibility measure (STOI)\footnote{https://github.com/mpariente/pystoi}. The STOI value for the additive noise is higher than the replacement noise, indicating that the target phrase is more recoverable when masked with additive noise.

\section{Discussion and Future Work}
We have discussed some of the current issues surrounding speech content privacy. While some of the scenarios for this capability will require slightly different solutions to be implemented, speech disentanglement is a promising overarching approach because it isolates speech content from other information in the speech signal. New progress in speech content privacy will be timely due to very recent advances in machine learning. It will have a large impact on society since privacy concerns influence how people adopt new voice technologies. One of the most important technical challenges will be processing privacy in real-time, and in a way that balances privacy needs with freedoms of expression. Allowing users a mechanism to adjust and control their content privacy settings will help create this necessary balance. Privacy controllability would allow users to have different privacy features in different settings, such as at home versus in public, or depending on who is nearby, or the type of voice device being used.

\section{Acknowledgements}
This work was partially supported by the EPSRC Centre for Doctoral Training in Data Science, funded by the UK Engineering and Physical Sciences Research Council (grant EP/L016427/1) and University of Edinburgh; and by a JST CREST Grant (JPMJCR18A6, VoicePersonae project), Japan.

\bibliographystyle{IEEEtran}
\bibliography{template}

\end{document}